\documentclass[aps,prb,preprint,showpacs,superscriptaddress,amsmath,amssymb]{revtex4}

\usepackage{graphicx}
\usepackage{dcolumn}
\usepackage{bm}

\begin{document}

\title{A Tight-Binding Grand Canonical Monte Carlo Study of the
  Catalytic Growth of Carbon Nanotubes}

\author{H. Amara} \affiliation{Laboratoire de Physique du Solide,
  Facult\'es Universitaires Notre-Dame de la Paix, 61 Rue de
  Bruxelles, 5000 Namur, Belgique} \affiliation{PCPM and CERMIN,
  Universit\'e Catholique de Louvain, Place Croix du Sud 1, 1348
  Louvain-la Neuve, Belgique}

\author{C. Bichara}
\affiliation{CRMCN-CNRS, Campus de Luminy,
Case 913, 13288 Marseille Cedex 09, France.}

\author{F. Ducastelle}
\affiliation{Laboratoire d'Etude des Microstructures,
ONERA-CNRS, BP 72, 92322
Ch\^atillon Cedex, France.}

\date{\today}

\begin{abstract}
  
  The nucleation of carbon nanotubes on small nickel clusters is
  studied using a tight binding model coupled to grand canonical Monte
  Carlo simulations. This technique closely follows the conditions of
  the synthesis of carbon nanotubes by chemical vapor deposition. The
  possible formation of a carbon cap on the catalyst particle is
  studied as a function of the carbon chemical potential, for
  particles of different size, either crystalline or disordered. We
  show that these parameters strongly influence the structure of the
  cap/particle interface which in turn will have a strong effect on
  the control of the structure of the nanotube. In particular, we
  discuss the presence of carbon on surface or in subsurface layers.

\end{abstract}


\maketitle

Accurately controlling the location, diameter, chirality and length of
single wall carbon nanotubes elaborated by chemical vapor deposition
(CVD) is one of the critical issues for an effective use of the unique
properties of these nanoobjects in many applications such as
electronic devices \cite{Dresselhaus2001}.  According to Nasibulin
\emph{et al.}\cite{Nasibulin_2005} and Bachilo \emph{et
  al.},\cite{Bachilo_2003} the diameter of the tubes produced by CVD
is related to the diameter of the catalyst particle from which they
originate. One important step is therefore to control the size
distribution and the location of the catalyst particle. Chirality may
result from either thermodynamic stability differences between the
initial tube caps \cite{Reich_chirality} or from difference in the
growth kinetics. \cite{Lolli}  The tube length can be limited by the
poisoning of the catalyst resulting from uncontrolled growth parameters.\\

The very large number of parameters controlling the CVD reaction makes
it necessary to rationalize the approach by an understanding of
nucleation and growth mechanisms at the atomic level.  Experimental
investigations at this level are usually done by Transmission Electron
Microscopy (TEM) performed \emph{post mortem} outside the CVD reactor.
Remarkable progress has recently been made in the \emph{in situ}
observation of the growth of nanotubes,\cite{Helweg, Zhu_small, Hofman_NL_2007} 
but the atomic resolution is not obtained under the actual growth conditions.\\

Computer simulation techniques are an alternative way to gain an
insight at the atomic level. However, they have to face the very
difficult challenge of requiring an accurate description of the
interatomic interactions, as typically provided by first principles
calculations, and to address the size and time scales relevant to the
experimental situation. Efficient catalyst particles for CVD reactions
are in the 1--10 nm diameter range (some thousands of atoms) and,
according to Lin \emph{et al.},\cite{Lin_Nanoletters_2006} the time
scale for the nucleation and growth of tubes is in the 10--100 seconds
range. Such scales are largely beyond the capabilities of Molecular
Dynamics (MD) techniques, even with the most simple interatomic
interaction models.  This means that only some elementary steps can be
studied, or that the growth conditions imposed in the simulations are
orders of magnitude too fast, resulting in very defective structures
as compared to the almost perfect tubes obtained experimentally.\\

Static \textit{ab initio} calculations have been used to calculate the
total energies of atomic configurations considered as representative
of the nucleation of carbon caps on a catalyst.\cite{Fan2003} Hofmann
\emph{et al.}  \cite{Hofmann2005} proposed a schematic growth model
and calculated the relevant activation barriers.  Abild-Pedersen
\emph{et al.}  \cite{Abild2006} showed that carbon surface or
subsurface diffusion on Ni correspond to almost equal diffusion
barriers. First principles Molecular Dynamics simulations were used to
study the root incorporation of C atoms on a Co catalyst
\cite{Gavillet2001} and the nucleation of a carbon cap on a Fe droplet
was described by Raty \emph{et al.}. \cite{Raty2005} As argued by
these latter authors the absence of diffusion inside the Fe particle
is due to its very small size (55 atoms). This is at odds with the
findings of Ding and Bolton \cite{Ding2006} who argued that highly
supersaturated carbon concentrations are required for nucleating
carbon islands on the Fe particle surface. This difference might be
due to the FeC interaction (empirical versus \emph{ab initio}) model
used or to the very short physical time of the first principles
simulations. The same group studied the roles of the particle size and
of a possible temperature gradient on the nanotube growth.\cite{Ding2004} 
Based on their empirical MD simulations, Shibuta and
Maruyama \cite{Shibuta2003} found hexagonal carbon networks formed
within a small Ni cluster of 108 atoms, in complete contradiction with
the \textit{ab initio} calculations of Zhang \emph{et al.}
\cite{Zhang2004} showing that, at 0 K, the most stable position for
one C atom on a small Ni$_{38}$ hexagonal cluster is adsorbed on the
surface rather than in the central interstitial site.  Studying the
growth of nanotubes on small (Ni$_{48}$ and Ni$_{80}$) clusters by
empirical MD, Zhao \emph{et al.}  \cite{Balbuena_Nanotechnology} found
almost the same mechanism as in our previous calculations on a semi
infinite system (slab geometry)\cite{Hakim2006} based on a tight
binding model. In both studies, four steps can be identified: C
dissolution, the formation of C chains on the surface followed by the
formation of $sp^{2}$ C sites that gradually detach from the catalyst surface.\\

Summarizing the computer simulations results, it seems that the
question of the presence of carbon dissolved in the catalyst remains
quite controversial. First principles static or dynamic calculations
tend to show that no C is dissolved in the very small particles
studied, while empirical simulations tend to find a larger amount of C
dissolved. Besides the questions of the accuracy of the interatomic
force model and of the time scale of the simulations, another aspect
is completely neglected in the MD calculations used. Molecular
Dynamics works by construction in a microcanonical or canonical
ensemble with a fixed number of particles. Growth sequences are
obtained either by adding particles ``by hand" or by putting the
catalyst particle in a box, with a fixed number of C atoms in a vapor
phase. Such a process completely ignores the chemical
potential gradient that is the thermodynamic driving force for the growth.\\

In this paper, using a carefully tested tight binding model we show
that the solubility of carbon and the possibility to grow tube embryos at
the surface of small Ni cluster critically depend on the carbon chemical
potential. Moreover, we show that, at 1000 K, the solubility of carbon
also depends on the structure of the small metallic cluster considered
: disordered, liquid-like clusters incorporate more C atoms, while
they tend to remain adsorbed on the surface of crystalline structures.\\

The tight binding model we used for C and Ni interactions has already
been used by Amara \emph{et al.} \cite{Hakim2006} and is described in
detail in reference \onlinecite{Amara2005}. The total energy is taken as a
sum of a band structure term and an empirical repulsive term. The
\textit{s} and \textit{p} electrons of C and \textit{d} electrons of
Ni are included.  Calculations are performed in the grand canonical
ensemble, with fixed volume, temperature (T), number of Ni atoms and C
chemical potential ($\mu_{C}$). In order to be efficiently implemented
in the grand canonical Monte Carlo (GCMC) code, the total energy has
to be taken as a sum of local terms, avoiding to recalculate the total
energy of the whole box at each step of the Monte Carlo process. This
is achieved by calculating local densities of electronic states using
the recursion method.\cite{Recursion} To keep the model as simple and
fast to compute as possible, we neglect the Ni \textit{s} electrons
and calculate only the first four moments of the local densities of
states. The local energy of each atom therefore depends only on the
positions and chemical identities of its first and second neighbors,
as defined by a cut off distance set at 3.20 \AA\ for first neighbors.\\

The GCMC calculations presented here compare the adsorption of carbon
atoms on small Ni clusters, either ``crystalline'', in their face
centered cubic equilibrium Wulff shape, with 201, 405, or 807 atoms,
or disordered, resulting from the quench of liquid droplets with 50 or
108 atoms. These calculations were done at 1000 K, a typical
temperature for the CVD synthesis of single wall nanotubes.  We
analyze the influence of the carbon chemical potential on the carbon
nanostructure grown on the Ni cluster, on the structure of the nickel
carbon interface and on the solubility, if any, of carbon in the Ni
cluster. The decomposition reaction of the carbon rich feedstock under
specific partial pressure conditions is not taken into account
explicitly in our calculations. However, the result of this reaction
is to provide atomic carbon with a given chemical potential that is
able to adsorb and diffuse on, or in, the nickel surface. As explained
in e.g. Lolli \emph{et al.},\cite{Lolli} this carbon chemical
potential is experimentally controlled through the thermochemistry of
the decomposition reaction. To achieve this in the computer
experiments, we use Monte Carlo simulations in the grand canonical
ensemble.\cite{Frenkel1996}\\

The grand canonical algorithm used consists in a series of Monte Carlo
macrosteps. Each macrostep randomly alternates displacement moves for
Ni or C atoms, attempts to incorporate carbon in a previously defined
active zone and attempts to remove existing carbon atoms. Within each
macrostep, we systematically performed four times the number of atom
attempted displacement steps. To incorporate carbon atoms in the
structure, 1000 attempts were made and ended as soon as one successful
incorporation has occurred. The number of attempted carbon removal
steps was equal to twice the number of C atoms present in the
structure. We typically performed up to 5000 macrosteps. We will see
below that this number of macrosteps is large enough to reach
equilibrium when the chemical potential conditions are such that no
carbon nanostructure grows. Of course, under growth conditions, no
equilibrium is ever reached and the number of adsorbed C atoms keeps
increasing. The maximum amplitude of the attempted displacements of Ni
or C atoms is adjusted during the run in order to have a rejection
rate around 50\%, ensuring an optimal use of the computer resources to
sample the configurational space of the system.\cite{Frenkel1996} On
average, the maximum amplitude of the displacements of Ni atoms turns
out to be twice as large as that of C atoms, meaning that the mobility
of Ni atoms is larger. In order to mimic the CVD process, the active
zone for inserting or extracting carbon atoms is defined as a region
of space at less than 3 \AA\ above and 3 \AA\ below the surface of the
Ni cluster.  Moreover, to avoid the encapsulation of the Ni cluster
with carbon and to account for the fact that, in the experiments, the
catalytic particles are supported, the active zone is limited to the
topmost (60\%) part of the cluster.  We stress that this is a very
crude approximation that neglects the modifications of the cluster
structure that can be induced by the substrate.\cite{Vervisch_2002}
In the case of the largest fcc Wulff-shaped clusters, this zone is
smaller: 50\% for the 405 atoms cluster and 40\% for the 807 atoms
cluster. A first reason for this is to limit the CPU time of the
calculations, since the fourth moment tight binding model for Ni and C
atoms is replaced by a much simpler second moment approximation model
for Ni atoms only, as soon as they are far enough from any C atom (i.
e.: at a distance larger than 6.4 \AA\ ). A second reason is that
these fractions of the clusters volume roughly correspond to the same
number of Ni atoms (151 for the 201 atom cluster, 231 for the 405
atoms one and 267 for the 807 atoms one), which will be used to
calculate the mole fraction of carbon atoms dissolved in the Ni structure.\\

To analyze the results of the calculations, we will have to
discriminate between carbon atoms that are outside the Ni cluster,
generally forming $sp$ (small chains) or $sp^{2}$ (layered) structures
and individual carbon atoms that are adsorbed on the surface or
incorporated in subsurface or bulk. A convenient way to do this is to
calculate the distance between the carbon atom of interest and its
nearest carbon neighbor. If this distance is smaller than 1.7 \AA\ ,
the atom is defined as an ``outer'' C atom, because C--C bond lengths in
$sp$ or $sp^{2}$ bonding states are smaller than this. If it is longer,
the carbon atom is either adsorbed on the Ni surface, with Ni
neighbors in the 1.85--1.95 \AA\ distance range, or incorporated in
interstitial sites: such carbon atoms are denoted as ``inner'' carbon
atoms in the following. This criterion stems from the analysis of the
carbon carbon pair correlation function that clearly shows distinctive
features for each type of atoms. It can be easily checked by a visual
inspection of the structures formed.\\

\begin{figure}[htbp!]
\includegraphics[width=16cm]{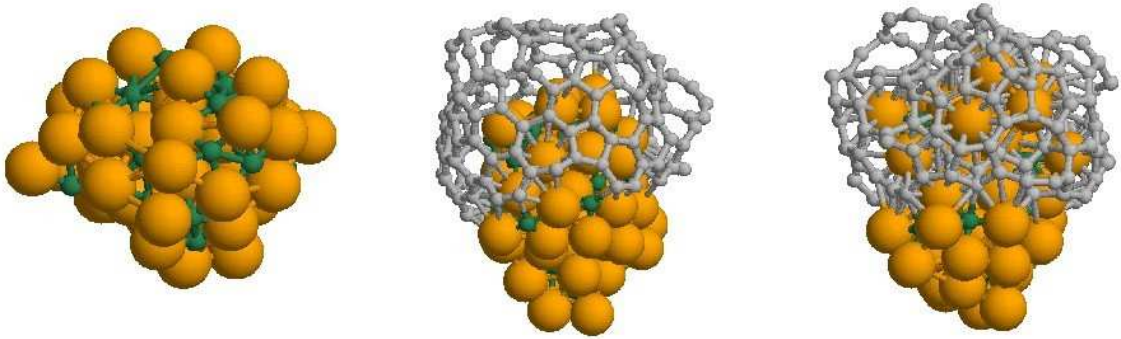}
\caption{Typical configurations resulting from GCMC simulations of the
  growth of C on a 50 Ni atom disordered droplet at 1000 K. Left panel,
  $\mu_{C}$ = -6.00eV/atom, central panel $\mu_{C}$ = -5.25 eV/atom,
  right panel $\mu_{C}$ = -4.50 eV/atom. 5000, 4000 and 2500 Monte
  Carlo macrosteps were performed, respectively. Nickel atoms are
  represented in orange, inner carbon atoms are in green and outer
  carbon atoms are in grey. See text for the definition of ``inner'' and
  ``outer'' carbon atoms.}
\label{Figure_Texas1}
\end{figure}

We first study the effect of the carbon chemical potential on the
structures formed outside the particle. Starting with the same
Ni$_{50}$ disordered droplet, we performed GCMC runs with $\mu_{C}$
varying between -6.00 and -4.50 eV/atom, by steps of 0.75 eV/atom. The
carbon chemical potential is referred to a fictitious ideal monoatomic
gas, explaining its values that are of the order of the cohesive
energies of the various carbon phases (e.g.: -7.41 eV/atom for a
graphene layer in our model). Figure \ref{Figure_Texas1} shows a
series of typical configurations obtained under these conditions. At
low chemical potential, no outer carbon structure is formed. Carbon
atoms land on the surface and tend to diffuse on the surface or
towards subsurface sites. They sometimes form carbon dimers, but no
twofold coordinated long chain structure is ever formed. An
equilibrium situation is quickly reached, with about 25 C atoms
adsorbed. On the contrary, when the chemical potential is much larger
(-4.5 eV/atom), a disordered and thick carbon layer is formed outside
the cluster. This amorphous layer is formed by carbon atoms with 3 or
4 carbon neighbors. This can be interpreted as the growth of an
amorphous carbon fiber. A much more interesting situation, as far as
the formation of a carbon nanotube is concerned, is observed at intermediate chemical potential values.\\

\begin{figure}[htbp!]
\includegraphics[width=17cm]{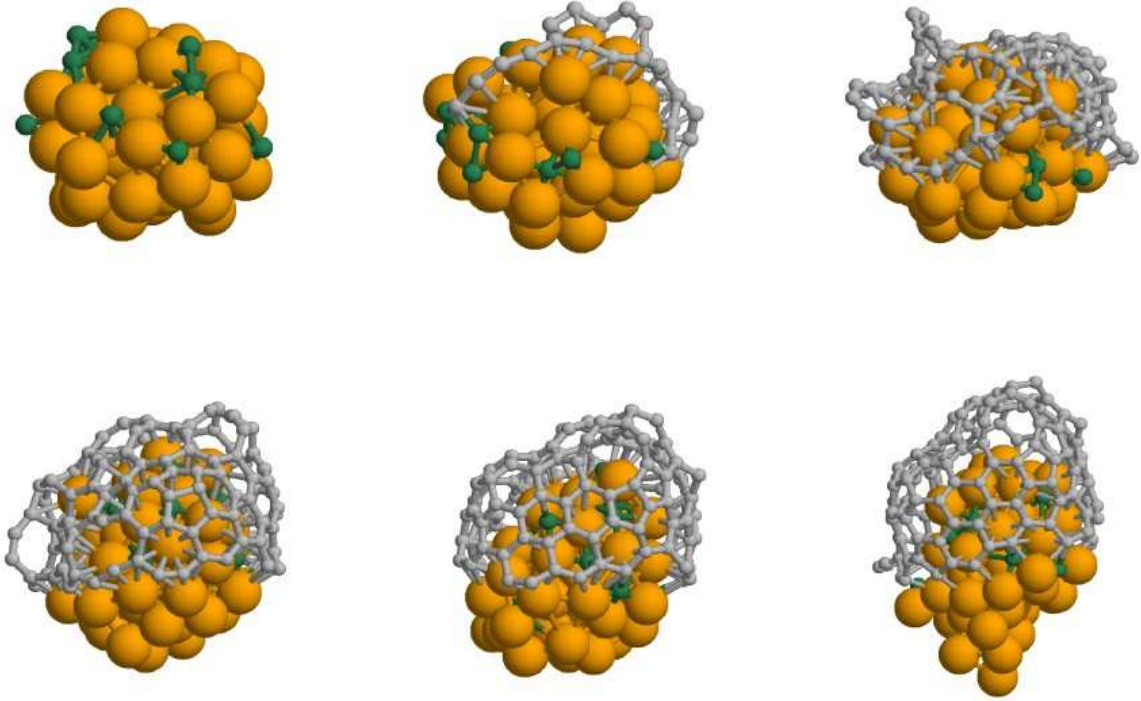}
\caption{Typical growth sequence during a GCMC simulation of the
  adsorption of C on a 50 Ni atom disordered droplet at 1000 K and
  $\mu_{C}$ = -5.25 eV/atom. From left to right and top to bottom: 1)
  carbon atoms or diatoms adsorbed; 2) chains growing; 3 and 4) cap
  formed; 5 and 6) cap detaches from the Ni droplet. Nickel atoms are
  represented in orange, inner carbon atoms are in green and outer
  carbon atoms are in grey. See text for the definition of ``inner'' and
  'outer'' carbon atoms.}
\label{Figure_Texas2}
\end{figure}

At $\mu_{C}$ = -5.25 eV/atom and 1000 K, a cap of carbon atoms is
formed. It encapsulates the topmost part of the Ni droplet. Figure
\ref{Figure_Texas2} presents a series of configurations characteristic
of the formation of the cap. At the very beginning of the GCMC run,
carbon atoms ``land'' on the available surface of the Ni particle. Then,
instead of diffusing towards subsurface sites as observed at lower
chemical potentials, they tend to form dimers and short chains at the
surface of the droplet. Because $\mu_{C}$ is larger, the flux of
incoming atoms is larger and C atoms quickly find another C to form a
C--C bond that stabilizes it on the surface. Although no time scale is
included in the grand canonical Monte Carlo scheme, we can see that
this step is very fast: in less than 50 Monte Carlo macrosteps,
carbon chains are already formed on the surface. As observed in our
previous calculations on Ni slabs,\cite{Hakim2006} the chains grow,
cross each other to form threefold coordinated $sp^{2}$ carbon sites
that act as nucleation centers to grow $sp^{2}$ layers. These $sp^{2}$
carbon atoms interact weakly with the Ni surface and the cap formed
can gradually detach from the droplet. The difference with
calculations performed on flat surfaces is that the curvature of the
Ni droplet induces a curvature of the carbon cap. The cap diameter and
curvature are determined by the Ni particle shape at the moment when
the cap is formed. However, since Ni atoms are more mobile than C
atoms, the shape of the Ni droplet can be modified later in the
process. As compared to the almost perfect graphene like layer, mostly
formed with hexagons, obtained on a flat Ni surface, the curvature and
growth conditions impose a larger number of defects such as
pentagons, heptagons,  etc.  The cap structure shown in figure
\ref{Figure_Texas1} is formed by 13 pentagons, 28 hexagons, 7
heptagons, 5 octogons and 7 9-membered rings. The cap is very
defective as compared to ideal nanotube terminations that require only
a small number of pentagons to be formed.\cite{Brinkmann_1999}  The
larger number of defects observed here results from the growth
conditions: the carbon chemical potential chosen to obtain realistic
structures in a reasonable, although long, CPU time is probably too
high. Under such conditions, the growth is too fast and the defects
formed have a very small
probability to be healed.\\

\begin{figure}[htbp!]
\includegraphics[width=17cm]{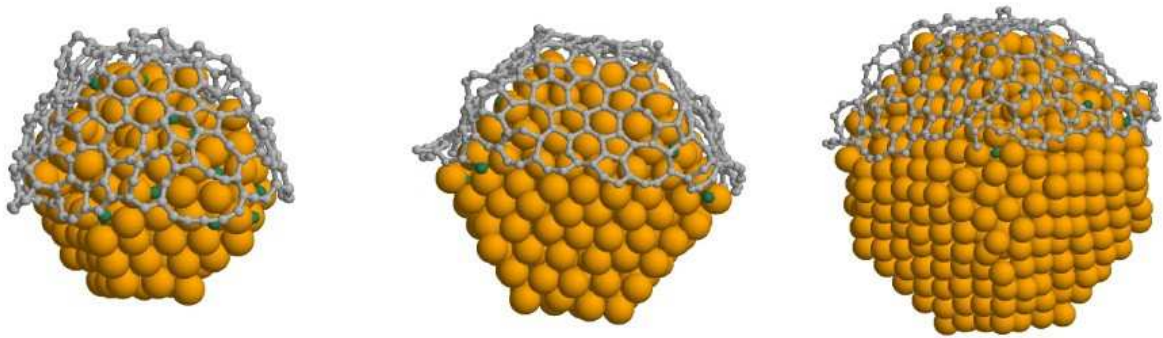}
\caption{Typical configurations resulting from GCMC simulations of the
  growth of C on nickel fcc Wulff-shaped clusters at 1000 K and
  $\mu_{C}$ = -5.25 eV/atom. Left: 201, center: 405 and right: 807
  Ni atoms. The active zones for the GCMC calculations were the
  topmost 60\%, 50\% and 40\% respectively (see text). Nickel atoms
  are represented in orange, ``inner'' carbon atoms are in green and
  ``outer'' carbon atoms are in grey. See text for the definition of
  ``inner'' and ``outer'' carbon atoms.}
\label{Figure_Texas3}
\end{figure}

The GCMC calculations presented above indicate that there is an
optimal carbon chemical potential to grow $sp^{2}$ carbon caps on a
very small Ni droplet. When the carbon chemical potential is large
enough, carbon nanostructures are formed outside the Ni droplet, while
some carbon atoms remain adsorbed on the surface, in close contact
with Ni atoms, or even diffuse inside the particle. The mole fraction
of these inner C atoms depends on the carbon chemical potential and on
the temperature, as indicated by comparing with our previous
calculations at 1500 K.\cite{Hakim2006} We note here that it also
depends on the particle size and on the structural state of the Ni
cluster. In order to analyze these latter points, we performed the
same calculations on a larger disordered droplet with 108 Ni atoms and
on face centered cubic clusters with a truncated octahedron shape that
minimizes their total (surface + volume) energy. These so called
Wulff-shaped clusters have only (111) and (100) facets. Three sizes
were considered: 201, 405 and 807 Ni atoms. At 1000 K and $\mu_{C}$ =
-5.25 eV/atom, the results are qualitatively the same, as far as the
cap formation is concerned. As shown in figure \ref{Figure_Texas3},
more or less defective carbon layers are formed on the cluster
surface. Their shape is imposed by the cluster geometry. In order to
allow them to detach from the surface, much longer runs should be
performed, as has been done for the Ni$_{50}$ disordered cluster. It
is interesting to notice that the crystalline structure of the cluster
is preserved, although the Ni atoms on the edges are less stable than
the others and can be strongly displaced, as shown on the right hand image in figure \ref{Figure_Texas3}.\\

\begin{figure}[htbp!]
\includegraphics[width=17cm]{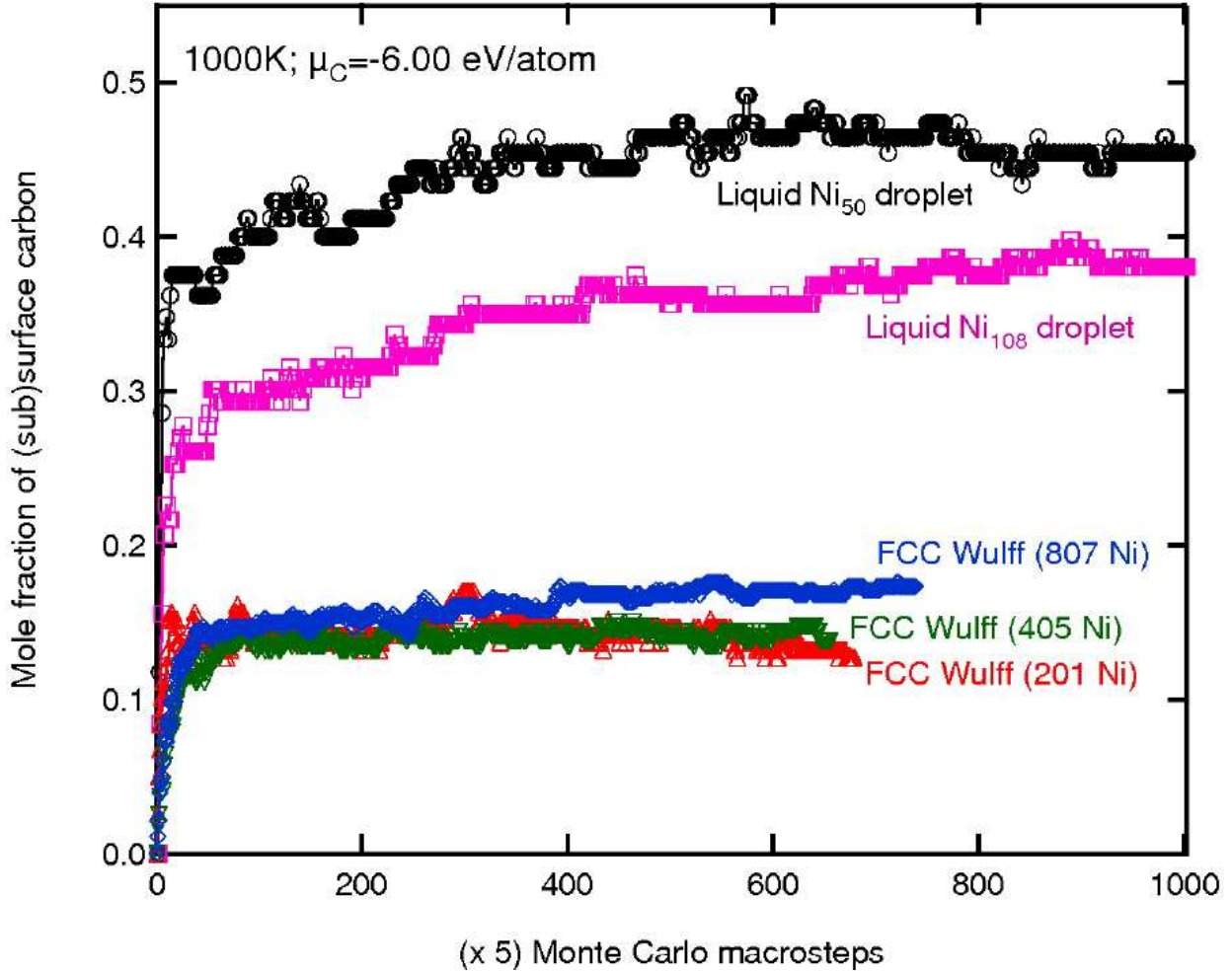}
\caption{Mole fractions of surface and subsurface carbon atoms,
  ``inner'' carbon atoms in the text, as a function of the number of
  outer Monte Carlo loops (macrosteps), calculated at 1000 K and
  $\mu_{C}$ = -6.00 eV/atom. Black circles: Ni$_{50}$ liquid droplet.
  Pink squares: Ni$_{108}$ liquid droplet. Red up-pointing triangles:
  fcc 201 Ni cluster. Green down-pointing triangle: fcc 405 Ni
  cluster. Blue diamond: fcc 807 Ni cluster.}
\label{Figure_Texas4}
\end{figure}

\begin{figure}[htbp!]
\includegraphics[width=17cm]{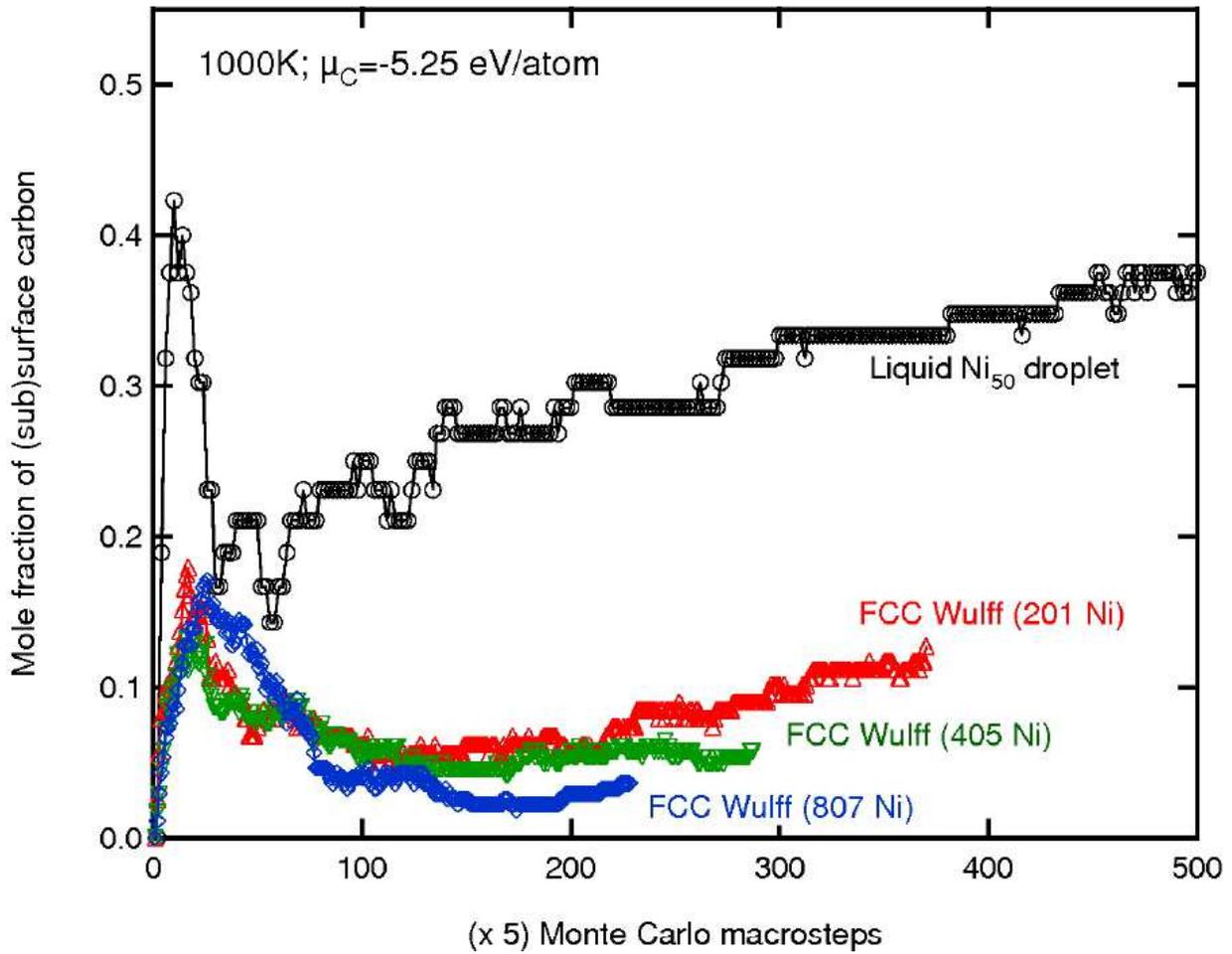}
\caption{Mole fractions of surface and subsurface carbon atoms,
  ``inner'' carbon atoms in the text, as a function of the number of
  outer Monte Carlo loops (macrosteps), calculated at 1000 K and
  $\mu_{C}$ = -5.25 eV/atom. Black circles: Ni$_{50}$ liquid droplet.
  Red up-pointing triangles: fcc 201 Ni cluster. Green
  down-pointing triangle: fcc 405 Ni cluster. Blue diamond:
  fcc 807 Ni cluster.}
\label{Figure_Texas5}
\end{figure}

A visual inspection of the structures obtained seems to indicate that
the number of dissolved C atoms is smaller in the case of crystalline
structures than in the disordered ones. This is confirmed by the
analysis of the mole fraction of inner carbon atoms as a function of
the number of Monte Carlo macrosteps, plotted in figures
\ref{Figure_Texas4} and \ref{Figure_Texas5}. For the disordered
structures, the mole fraction of inner C atoms is calculated with
respect to the total number of Ni atoms (respectively 50 and 108),
because the particles are small and C atoms tend to diffuse in the
whole structure. As mentioned above, for the fcc clusters, the mole
fraction of inner C atoms is calculated with respect to a smaller
number of Ni atoms. These different ways of calculating the mole
fractions tend to minimize the differences between the carbon
solubilities in crystalline or disordered clusters. At 1000 K and
$\mu_{C}$ = -6.00 eV/atom (see figure \ref{Figure_Texas4}), the
difference is quite clear, with about 40\% C adsorbed on, or dissolved
in, the disordered cluster and about 15\% C atoms, mostly adsorbed on
the surface or subsurface sites of the crystalline clusters. The same
holds for the runs at $\mu_{C}$ = -5.25 eV/atom. However, a striking
difference can be noticed between the two situations. At $\mu_{C}$ =
-6.00 eV/atom the mole fraction of inner carbon atoms gently rises
during the course of the Monte Carlo run to reach an equilibrium
value. At $\mu_{C}$ = -5.25 eV/atom, the number of carbon atoms
adsorbed on the cluster surface increases very rapidly at the
beginning of the process, then decreases and rises again. During this
process the total number of carbon atoms constantly increases. The
sharp peak observed in the inner carbon concentration at the beginning
of the adsorption process corresponds to the adsorption of individual
C atoms on the cluster surface. As soon as the concentration of
surface C atoms is large enough, they tend to form chains and then
$sp^{2}$ structures that detach from the surface, explaining the
depletion of C at the surface. At lower chemical potential, this
surface concentration threshold is never reached and the C atoms
gradually diffuse towards subsurface or bulk interstitial
sites.\\

Summarizing this paper, we study the early stages of the formation of
carbon nanostructures on small nickel clusters. The GCMC method used
enables us to emphasize the critical role of the carbon chemical
potential on the resulting structure. We show that an optimal chemical
potential value exists to nucleate nanotube caps. Because of the short
time scale spanned by the computer simulations as compared to the
hundreds of seconds of the experiments, the cap structures obtained
are much more defective than what is assumed for the experimental
ones. The coarse sampling of the chemical potential values by steps of
0.75 eV/atom also plays a role: less defective caps could probably be
grown with $\mu_{C}$ closer to -6.00 eV/atom, but this would require
longer simulation times. Under suitable carbon chemical potential and
temperature conditions, the early stages of the growth are quite
similar to what has been observed on flat surfaces,\cite{Hakim2006} 
with the difference that the curvature is imposed by the catalyst
particle's size at the moment of the nucleation. The atomistic growth
mechanism is more or less in agreement with most of the previous
computer simulation studies.\cite{Raty2005,Ding2006,Ding2004,Balbuena_Nanotechnology}
However these authors do not take into account the carbon chemical 
potential that plays a critical role on the outer nanostructure formed, 
but also on the surface state of the catalyst. In addition, we show that 
the presence of carbon in surface, or interstitial subsurface or bulk sites
 depends on a second parameter that is the structure of the catalyst particle. 
As in the case of the bulk phase diagram that exhibits a low solubility of carbon in
crystalline nickel, but a larger one in the liquid phase, we show that
the carbon solubility is larger in disordered clusters. Our study is
not yet complete since we should compare liquid and solid structures
with the same number of atoms. This work is in progress.  However,
this is an important finding because the surface state of the catalyst
is of fundamental importance if one wishes to understand the origin of
the chiral selectivity that has been reported for instance by Bachilo
\emph{et al.}.\cite{Bachilo_2003}\\

\begin{acknowledgments}
  Part of this work is supported by the Belgian Program on Interuniversity
  Attraction Poles (PAI6) on ``Quantum Effects in Clusters and
  Nanowires''.  H. A.  acknowledges the use of the Interuniversity
  Namur Scientific Computing Facility (Namur-ISCF), a common project
  between FNRS, SUN Microsystems, and Les Facult\'es Universitaires
  Notre-Dame de la Paix (FUNDP) and the LEMAITRE computer of UCL.
\end{acknowledgments}

\end{document}